\documentclass[journal]{IEEEtran}

\usepackage{cite}
\usepackage{amsmath,amssymb,amsfonts}
\usepackage{algorithmic}
\usepackage{graphicx}
\usepackage{textcomp}

\usepackage{url}
\usepackage{float}
\usepackage{nccmath}
\usepackage{adjustbox}
\usepackage{multirow}
\usepackage{eurosym}
\usepackage{mathtools}
\usepackage{textcomp}
\usepackage{verbatim}
\usepackage{soul} 
\usepackage{enumitem}
\usepackage{physics}
\usepackage{dsfont}
\usepackage{color}
\usepackage{balance}
\usepackage{subfigure}
\usepackage{nccmath}
\usepackage{algorithm}
\usepackage{array}
\usepackage{stfloats}

\def\BibTeX{{\rm B\kern-.05em{\sc i\kern-.025em b}\kern-.08em
    T\kern-.1667em\lower.7ex\hbox{E}\kern-.125emX}}

\begin{document}

\title{Efficient Variational Quantum Algorithms for the Generalized Assignment Problem
}
\author{
Carlo Mastroianni,
Francesco Plastina,
Jacopo Settino, 
and Andrea Vinci 
\thanks{Carlo Mastroianni and Andrea Vinci are with the National Research Council of Italy - Institute for High-Performance Computing and Networking (ICAR-CNR), Via P. Bucci, 8/9 C, 87036 Rende (CS), Italy
(email: carlo.mastroianni@icar.cnr.it, andrea.vinci@icar.cnr.it)}%
\thanks{Francesco Plastina and Jacopo Settino are with the Dip. Fisica, Universit\`a della Calabria, Arcavacata di Rende (CS), Italy, (e-mail: francesco.plastina@fis.unical.it, jacopo.settino@unical.it) and also with INFN - Gruppo Collegato di Cosenza }%
\thanks{Corresponding author: Andrea Vinci (email: andrea.vinci@icar.cnr.it).}
}

\maketitle

\begin{abstract}
Quantum algorithms offer a compelling new avenue for addressing difficult NP-complete optimization problems, such as the Generalized Assignment Problem (GAP). Given the operational constraints of contemporary Noisy Intermediate-Scale Quantum (NISQ) devices, hybrid quantum-classical approaches, specifically Variational Quantum Algorithms (VQAs) like the Variational Quantum Eigensolver (VQE), 
promises to be effective approaches to solve real-world optimization problems. This paper proposes an approach, named VQGAP, designed to efficiently solve the GAP by optimizing quantum resources and reducing the required parametrized quantum circuit width with respect to standard VQE. The main idea driving our proposal is to decouple the qubits
of ansatz circuits from the binary variables of the General Assignment Problem, by providing encoding/decoding functions
transforming the solutions generated by ansatze in the limited
quantum space in feasible solutions in the problem variables
space, by exploiting the constraints of the problem.
Preliminary results, obtained through both noiseless and noisy simulations, indicate that VQGAP exhibits performance and behavior very similar to VQE, while effectively reducing the number of qubits and circuit depth.





\end{abstract}


\section{Introduction}
\label{secIntro}

Quantum algorithms provide a promising new approach to tackling complex optimization problems\cite{VQA, VQA-IEEE-TQE}—including examples like Max-Cut\cite{maxcut}, Max-Sat  \cite{maxsat}, and various routing problems \cite{routing-IEEE-TQE}. The methodology often stems from the adiabatic paradigm, which outlines how $\text{NP}$-complete problems can be mapped onto an Ising model \cite{IsingNP}. Within this framework, the solution is represented as the minimal energy state of the system's Hamiltonian, and the quantum computer drives the system towards this state. This principle has inspired practical algorithms executable on available gate-based quantum computers. 

The operational constraints of contemporary Noisy Intermediate-Scale Quantum (NISQ) devices \cite{preskillNisq, NISQ-IEEE-TQE} necessitate the adoption of hybrid quantum-classical algorithms. This approach, exemplified by Variational Quantum Algorithms (VQAs)\cite{biamonte2017,dunjko2016, QuantumOptimization2018}, leverages classical optimization routines to iteratively tune the parameters driving the quantum computation. The goal is to identify an optimal solution with a significant computational speed-up. VQAs offer key advantages: they simplify the formulation of problem-specific quantum circuits, as opposed to the often non-intuitive nature of purely quantum algorithms. Critically, by delegating parameter optimization to the classical domain, VQAs are presumed to mitigate hardware demands, requiring minimal qubit counts and shallow quantum circuit depth.

Our prior research, for example, successfully applied these techniques to optimize energy exchange protocols within a prosumer network\cite{IEEESmartGrid} and process allocation in edge/cloud computing networks \cite{QuantumEdgeCloud}, in particular by exploiting the Quantum Approximate Optimization Algorithm (QAOA) \cite{QAOA} and the Variational Quantum Eigensolver (VQE) \cite{VQE}. 

QAOA and VQE encompass two distinct strategies for encoding an optimization problem into a quantum circuit, often called an ansatz.

QAOA (Quantum Approximate Optimization Algorithm), inspired by the adiabatic theorem, uses an alternating sequence of two Hamiltonians—one representing the problem—to drive the system toward the problem's minimum energy state (the solution). Its key strength is a guarantee of convergence as the circuit depth (number of repetitions) increases. However, a deep circuit makes it very susceptible to noise, and it fails to reduce the exponentially growing search space of the problem.

VQE (Variational Quantum Eigensolver) offers a more flexible approach where the circuit designer can customize the ansatz to the specific problem. This flexibility, however, means it lacks the convergence guarantees of QAOA.

This latter algorithm, in particular, has been demonstrated to be effectively exploitable during the current NISQ era, as it provides a significant reduction in quantum circuit depth by carefully designing parametrized quantum circuits tailored for solving specific optimization problems, compared to QAOA.

Moving from the Variational Quantum Eigensolver, this paper aims at proposing an approach for efficiently solving the well-known Generalized Assignment Problem by exploiting gate-based quantum computing, by providing a solution for reducing the width of the required parametrized quantum circuit with respect to the VQE. This aspect is crucial when trying to solve real-world problems with the current available quantum hardware, since the number of available qubits is still very limited.

The main idea driving our proposal is to decouple the qubits of ansatz circuits from the binary variables of the General Assignment Problem, by providing encoding/decoding functions transforming the solutions generated by ansatze in the limited quantum space in feasible solutions in the problem variables space, by exploiting the constraints of the problem.

The rest of the paper is organized as follows: Section \ref{sec:modeling} introduces the formulation of the General Assignment Problem and describes how GAP can be re-formulated in a an quadratic uncostrained problem to be solved with VQE; Section \ref{sec:vqo} summarizes how the VQE works, and presents the main contribution of the paper, namely, VQGAP and its variation VQGAPe; Section \ref{sec:results} present and discuss a set of preliminary results for assessing the efficacy of the proposed approach. Finally, Section \ref{sec:conclusions} concludes the paper and discusses some perspectives for future research work.














\section{Modelling of the Generalized Assignment Problem}
\label{sec:modeling}





In this section, we introduce and present the well-known formulation of the General Assignment Problem, 
and describe the transformation needed to tackle it using the Variational Quantum Eigensolver algorithm.
\subsection{Formulation of the General Assignment Problem}
\label{SubSecProblems}

Given a set of tasks $\mathcal{T}=\{1, ..., T\}$ and a set of agents $\mathcal{A}=\{1, ..., A\}$, the problem is to optimally assign tasks to agents. Two kinds of constraints must be matched: each agent $j$ has a limited budget $B_j$, and each task can be assigned to a single agent, or can remain unassigned. 



A set of binary variables $x_{ij}$ are defined, which take the value $1$ if the task $i$ is assigned to the agent $j$, and $0$ otherwise. A task $i$ can remain unassigned. The objective is to assign the tasks to agents by maximizing the overall profit, while matching the budget constraints of the agents. The profit and the constraints can be written as:

\begin{align}
     \label{eq:objFunction}
     \max & \sum_{ i \in \mathcal{T},j \in \mathcal{A}} p_{ij}x_{ij}  \\
     \label{eq:vincoliProcesso}
    & \sum_{j \in \mathcal{A}} x_{ij} \leq 1 , ~~ & \forall i \in \mathcal{T} \\ 
    \label{eq:vincoliNodo}
    & \sum_{i \in \mathcal{T}} w_{ij}x_{ij} \leq B_{j} ,~~~ & \forall j \in \mathcal{A} \\
    & x_{i,j} \in \{0,1\},
    ~~~ & \forall i \in \mathcal{A} \land \forall j \in \mathcal{T}
\end{align}

where each task $i \in \mathcal{T}$ is assigned a value $p_{ij}$, which is the profit deriving by assignment the task $i$ to the agent $j \in \mathcal{A}$, and a weight $w_{ij}$, which represents the amount of resources needed by agent $j$ to carry out task $i$. Each agent $j$ has a resource budget $B_j \in \mathbb{Z}^+$ .



The number of binary variables $x_{ij}$ is equal to $T \cdot A$. Our approach, later discussed in section \ref{sec:vqo} needs the transformation of inequalities into equations. In order to perform this transformation, a number of slack variables must be added. In particular, for each task $i$, one slack binary variable $s_i$ is needed to specify whether the task is actually assigned to an agent. Accordingly, by adding $T$ slack variables, constraints (\ref{eq:vincoliProcesso}) are reformulated as:

\begin{equation}
\label{eq:residualP}
\sum_{j \in \mathcal{A}} x_{ij} + s_i = 1
\end{equation}

Furthermore, to convert the inequalities (\ref{eq:vincoliNodo}) into equations, for each agent $j$, we add a slack variable $r_j \ge 0$, that represent the residual budget of the agent $j$, i.e., the budget capacity that remains not assigned to any task. The constraints (\ref{eq:vincoliNodo}) become:

\begin{equation}
\label{eq:residualB}
     \sum_{i \in \mathcal{T}} w_{ij}x_{ij} + r_j = B_j 
\end{equation}

Thus, the problem is formulated in terms of: (i)  $T \cdot (A+1)$ binary variables, where this number is the sum of the $T \cdot A$ assignment variables $x_{ij}$ and the $T$ slack variables $s_i$, and (ii) $A$ real variables, which are the residual slack variables $r_j$.

It is worth noting that in VQE algorithms, all the variables need to be mapped into qubits, and thus the residual capacity must be reformulated into a number of binary slack variables. Without loss of generality, we assume that both the budget values $B_j$ and the weight values $w_{ij}$ are integers, and thus, possible values of residual $r_{j}$ are integers too. In order to express all the possible residual budget values $r_{j}$, the constraints could be reformulated as follows:


\begin{align}
\label{eq:ineq_to_eq}
    & \sum_{i \in \mathcal{T}} w_{ij}x_{ij} + \sum_{k=1}^{\lceil \log_2(B_j+1) \rceil} 2^{(k-1)} b_{jk} = B_j
\end{align}
where we have expressed the value of the residual $r_j$ with $\lceil \log_2 (B_j+1) \rceil$ slack binary variables, denoted as $b_{jk}$. In this case, the problem is formulated in terms of a number of binary variables equal to:

\begin{equation}
\label{eq:nQubits}
    Q = T \cdot (A + 1) + 
    \sum_{j \in \mathcal{A}}
    \lceil \log_2(B_j+1) \rceil
\end{equation}

In VQE, all the binary variables need to be mapped to qubits, thus $Q$ is also the number of required qubits. 

In this paper, we propose an alternative strategy through which the residual variables do not need to be mapped to qubits but can be managed in the classical portion of the hybrid quantum optimization algorithm. By using this strategy, the required number of qubits is limited to $Q = T \cdot (A + 1)$.


%


\subsection{Problem Mapping for Quantum Optimization}
\label{subsecProblemMapping}

The described problem with constraints is equivalent to an unconstrained optimization problem in which an extended objective function incorporates the constraints in the form of penalties. The extended objective function is defined as:

\begin{multline} \label{eq:extObj}
     \min \bigg( - \sum_{i \in \mathcal{T},j \in \mathcal{A}}p_{ij}x_{ij} 
       \\
      + C \cdot \sum_{i \in \mathcal{T}}\Big(
      1 - \sum_{j \in \mathcal{A}} x_{ij} + s_i\Big)^2  \\
      + C \cdot \sum_{j \in \mathcal{A}}\Big(
        B_j - \sum_{i \in \mathcal{T}} w_ix_{ij} + r_j
      \Big)^2
      \bigg)
\end{multline}

where the value of the constant $C$ is defined as:

\begin{equation}
C = 1 + \sum_{ i \in \mathcal{T},j \in \mathcal{A}}p_{ij}
\label{eq:penalty}
\end{equation}

The maximization problem has been converted to a minimization problem, and each constraint has been transformed into a penalty, which is equal to 0 only when the constraint is satisfied by the values of the variables. When the constraint is not satisfied, the value of the penalty is equal to or larger than $C$. Since the value of $C$ is defined to be larger than the maximum possible value of the first term of Eq. (\ref{eq:extObj}),
the violation of even a single constraint cannot be compensated by the minimization of the first term. This ensures that the solution that minimizes the objective function is obtained with values of the variables that satisfy all the constraints.

\section{Efficient Variational Quantum Optimization for GAP}
\label{sec:vqo}

The optimization problem that we have just introduced can be tackled with gate-based quantum computing. After recalling the standard procedure, based on VQE, we present in this Section our approach, that we tamed VQGAP, and a further improved form of it, the VQGAPe approach. 

\subsection{VQE approach}
\label{sec:vqe-intro}
\begin{figure}[!tb]
\centering
\includegraphics[width=0.9\columnwidth, angle=0]{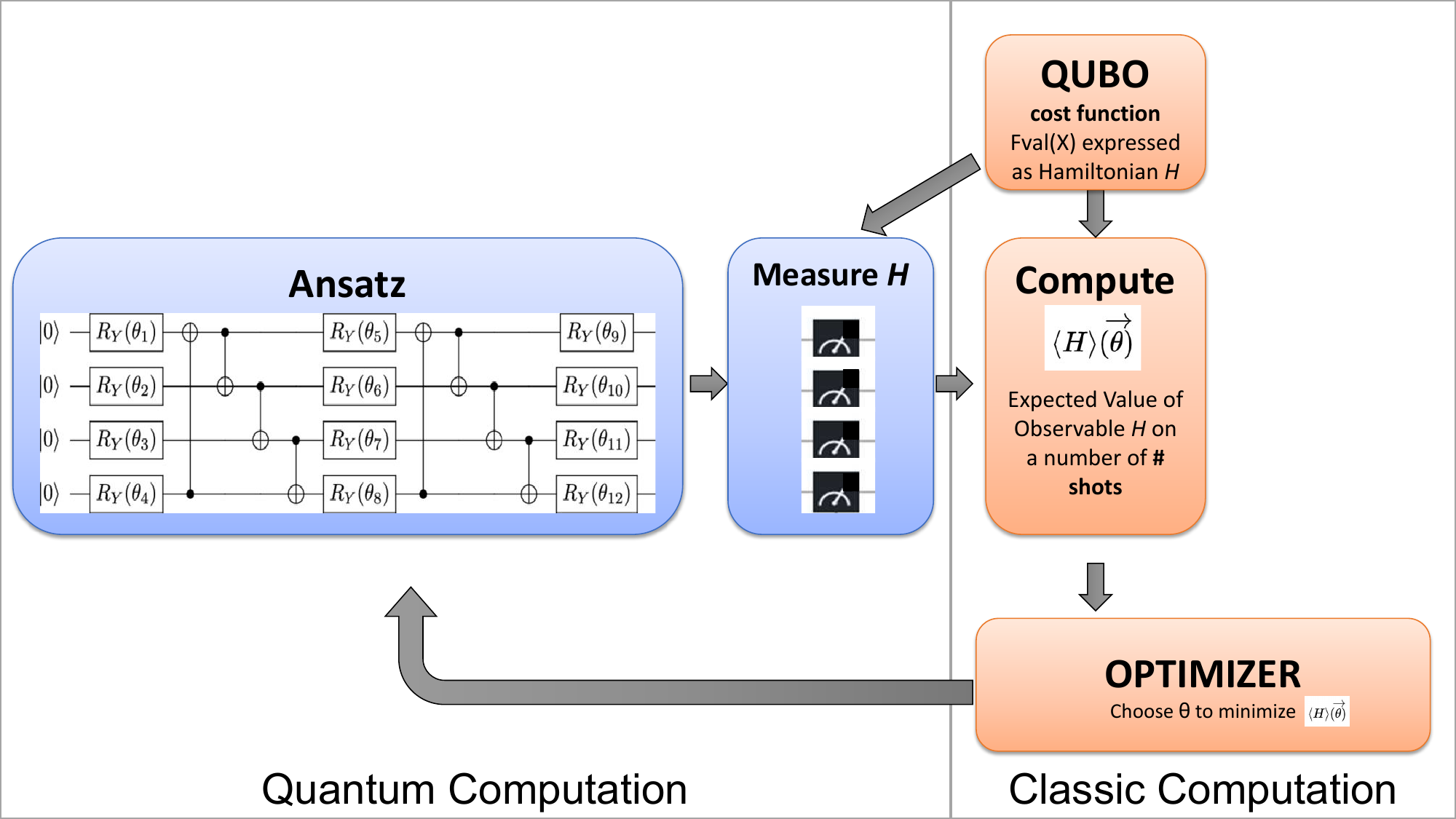}
\caption{Sketch of the VQE algorithm, where the execution of the Quantum circuit, structured according to the Ansatz, is followed by a series of measurements, dictated by the form of the QUBO problem. The Average of the Hamiltonian (i.e., our Cost Function) is then evaluated through the measurements, and its value serves as input to the Optimizer in order to change the parameters of the variational circuit.}
\label{fig:VQE}
\end{figure}

The VQE algorithm (see Figure \ref{fig:VQE}) requires expressing the optimization problem as a Hamiltonian Observable $H$, and a parametrized quantum circuit $U_{var}(\vec{\theta})$, referred to as ansatz, where $\vec\theta$ is a vector of parameters of the ansatz. VQE solves the problem by searching the parameters $\vec\theta$ that minimize the expectation value of the observable $H$, taken on the state obtained after the variational Unitary, $U_{var}(\vec\theta)$, has been run on the circuit. The dependence of $U_{var}$ on $\vec\theta$ is defined by the chosen Ansatz. In formulae:

\begin{equation} \label{eq:vqe-min}
\begin{split}
& \min_{\vec\theta} \bra{\phi(\vec\theta)}H\ket{\phi(\vec\theta)} \\
& \ket{\phi(\vec\theta)}= U_{var}(\vec\theta)\ket{0}
\end{split}
\end{equation}

With VQE, it is necessary to map all the problem variables to qubits, therefore the width of the quantum circuit is $Q$, as resulting from expression (\ref{eq:nQubits}). This requires the preliminary reformulation of the optimization problem into the form of an Ising problem. 

In particular, for every binary variable $\tilde{x}_i, i=1 \cdots Q$, with $i$ globally running on all the binary (including slack) variables, a corresponding discrete variable $z_i$ is defined with the substitution:

\begin{equation}
\label{eq:substitutionx}    
   \tilde{x}_i =\frac{1-z_i}{2}, ~~~ i = 1, ..., Q 
\end{equation}

After the substitution, the extended objective function is rewritten as a sum of terms $z_i$ and $z_i \cdot z_j$, and an Ising problem is obtained \cite{ising2000}, formulated as:

\begin{equation}
\label{eq:ising}
	min \ \bigg( {\sum_{i=1}^{Q}{h_i \cdot z_i} - \sum_{i=1}^{Q}\sum_{j=1}^{i-1}{J_{ij} \cdot z_i \cdot z_j}} \bigg)
\end{equation}
\\
where $h_i$ and $J_{ij}$ are real constants obtained after applying the substitutions.

In order to leverage the VQE algorithm, each variable is associated with one of the qubits of a quantum register. In particular, $z_i$ is given by the outcome of the measurement of the so-called $\textbf{Z}$ observable, performed on the $i$-th qubit at the end of the algorithm. According to quantum mechanics, the measurement has, indeed, the two possible outcomes $+1$ and $-1$, which are the two eigenvalues of  $\textbf{Z}$. Correspondingly, after the measurement, the state of each qubit collapses into one of the two logical states, denoted (using Dirac notation) by $\ket 0 = [1,0]^T$ and $\ket 1= [0,1]^T$. These are the eigenstates of the $\textbf{Z}$ operator, which
can expressed, in the logical basis, as the third Pauli Matrix:

\[
\textbf{Z} = \begin{bmatrix}
    1  & 0  \\
    0  & -1 \\
\end{bmatrix}
\]

Then, the Ising problem (\ref{eq:ising}) is mapped into a diagonal Hamiltonian operator, built with sums and tensor products (i.e., Kronecker products) of two basic one-qubit operators, the identity \textbf{I} and the Pauli operator \textbf{Z}. For each term in (\ref{eq:ising}), the operator $\textbf{Z}_i$ substitutes the variable $z_i$, and the identity operator $\textbf{I}_i$ is assumed to be inserted for each variable $z_i$ that does not appear explicitly. Moreover, the multiplications between two $z$ variables are substituted by the tensor products between the corresponding \textbf{Z} operators. For example, with $Q=4$, the term $z_2$$\cdot$$z_3$becomes $\textbf{I}_1 \otimes \textbf{Z}_2 \otimes \textbf{Z}_3 \otimes \textbf{I}_4$ or, more succinctly, $\textbf{I}_1 \textbf{Z}_2 \textbf{Z}_3 \textbf{I}_4$ or, even more briefly, $\textbf{Z}_2 \textbf{Z}_3$, where the identity operators are implicit.  With these rules, the Hamiltonian operator that corresponds to expression (\ref{eq:ising}) is:

\begin{equation}
\label{eq:IsingHamiltonian}
	\textbf{H} = {\sum_{i=1}^{Q}{h_i \cdot \textbf{Z}_i} - \sum_{i=1}^{Q}\sum_{j=1}^{i-1}{J_{ij} \cdot \textbf{Z}_i \otimes \textbf{Z}_j}}
\end{equation}

Now, the problem is to find the minimum eigenvector(s) of the operator (\ref{eq:IsingHamiltonian}), which corresponds to finding the string of values of $z$ variables that minimizes the Ising expression (\ref{eq:ising}). 

The VQE algorithm can be used to explore the Hilbert space and search for the ground state of the Ising Hamiltonian operator.

\begin{figure}[!tb]
    \centering
    \includegraphics[width=0.95\columnwidth]{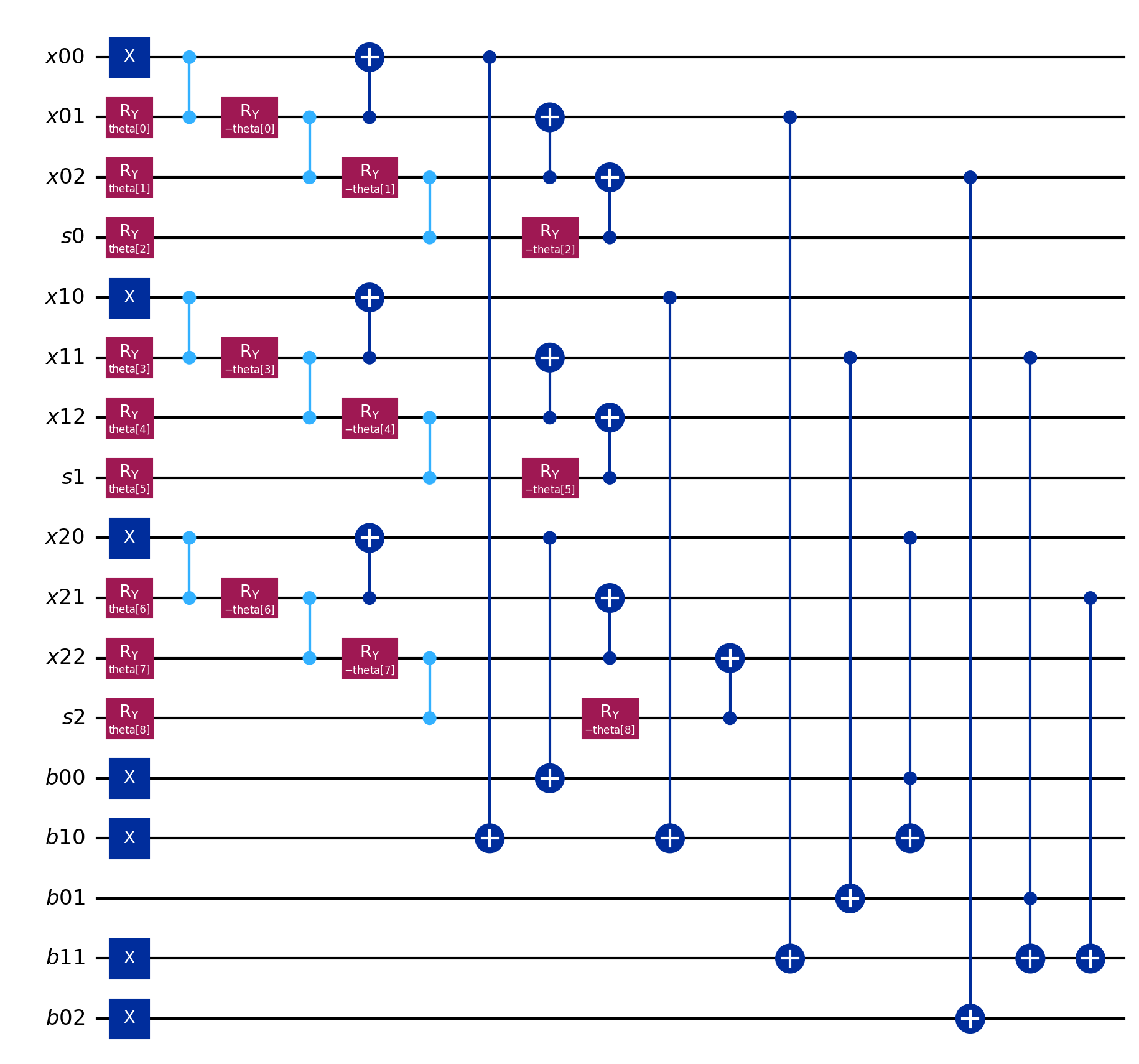}
    \caption{Reference ansatz for VQE. The state of the slack qubits is determined by the values of the assignment qubits. First, slack qubits are set to the capacities of respective agents (here $B_0=3$, $B_1=2$, $B_2=1$), then the task weights ($w_1$, $w_2$, and $w_3$) are subtracted if the related task is assigned to the agents.}
    \label{fig:ansatz_withslack_bounded_slack}
\end{figure}

For the GAP problem, we chose as a reference ansatz the one introduced in \cite{IEEE-TQE-2024}, and depicted in Fig. \ref{fig:ansatz_withslack_bounded_slack} for a GAP problem with $T=3$, $A=2$, $\{B_{j}\}=\{3,2\}$. This ansatz exploits two strategies: (i) it guarantees that each measurement matches the constraints in (Eq. \ref{eq:vincoliProcesso}), and (ii), given the assignment of tasks to agents (qubits $x_{ij}$), it computes the values of slack variables univocally, matching the constraints expressed in Eq \ref{eq:ineq_to_eq} whenever is possible (qubits $b_{kj}$).
Notice that this computation is performed in superposition on each state defined by the assignment qubits $x_{ij}$.
With this latter ansatz, the number of tunable parameters is equal to $\Theta = T \cdot A$. The slack qubits $b_{ji}$ represent the residual capacities of nodes; therefore, the gates on these qubits are used, first, to set the nominal capacities of the nodes, expressed as binary numbers, then to subtract from such values the weights of the processes that are assigned to the nodes.

For this ansatz, the number of two-qubits gates $G_2$ and its depth in terms of two-qubits gates $D_2$ is reported in the following:

$\mathcal{O}(G_2) = T\cdot (\sum_{j \in \mathcal{A}}(\lceil \log_2(B_j+1) \rceil)^3$      

$\mathcal{O}(D_2) = T \cdot (max_{j \in \mathcal{A}} \lceil log_2(B_j+1) \rceil)^2$

For the general case, the order of magnitude of $G_2$ and $D_2$, is polynomial in the number of tasks and in the capacity of agents, as reported in \cite{IEEE-TQE-2024}.




\subsection{VQGAP approach}
\label{sec:VQGAP}


Differently from VQE, which relies on evaluating and minimizing the expected value of the Hamiltonian observable (see Eq. \ref{eq:IsingHamiltonian}) which expresses the QUBO evaluation function defined in Eq. \ref{eq:extObj}, the proposed VQGAP aims at decoupling the ansatz and measurements from the computation of the expected value of the objective function, which becomes the target of a classical optimization module.


With VQGAP, each assignment binary variable $x_{ij}$ is mapped to a different qubit. A variational ansatz circuit is executed, after which each qubit is measured on the computation basis: the result of such measurement ($0$ or $1$, with respective eigenstates $\ket{0}$ or $\ket{1}$) is assigned to the corresponding binary variable. Given the assignment pattern, it is now possible to compute the values of the residual budgets $r_j$, according to:
\begin{equation}
\label{eq:residual-vqgap}
      r_j=\lvert B_j - \sum_{i \in \mathcal{T}} w_{ij}x_{ij} \rvert
\end{equation}
that produces values of $r_j$ that are consistent with Eq.~\ref{eq:residualB} for solutions that are feasible with respect to capacity constraints expressed in Eq.~\ref{eq:vincoliNodo}, otherwise the values of $r_j$ will not fulfill Eq.~\ref{eq:residualB}, thus activating the related penalty expressed in Eq.~\ref{eq:extObj}.

\begin{figure}[!tb]
\centering
\includegraphics[width=0.9\columnwidth, angle=0]{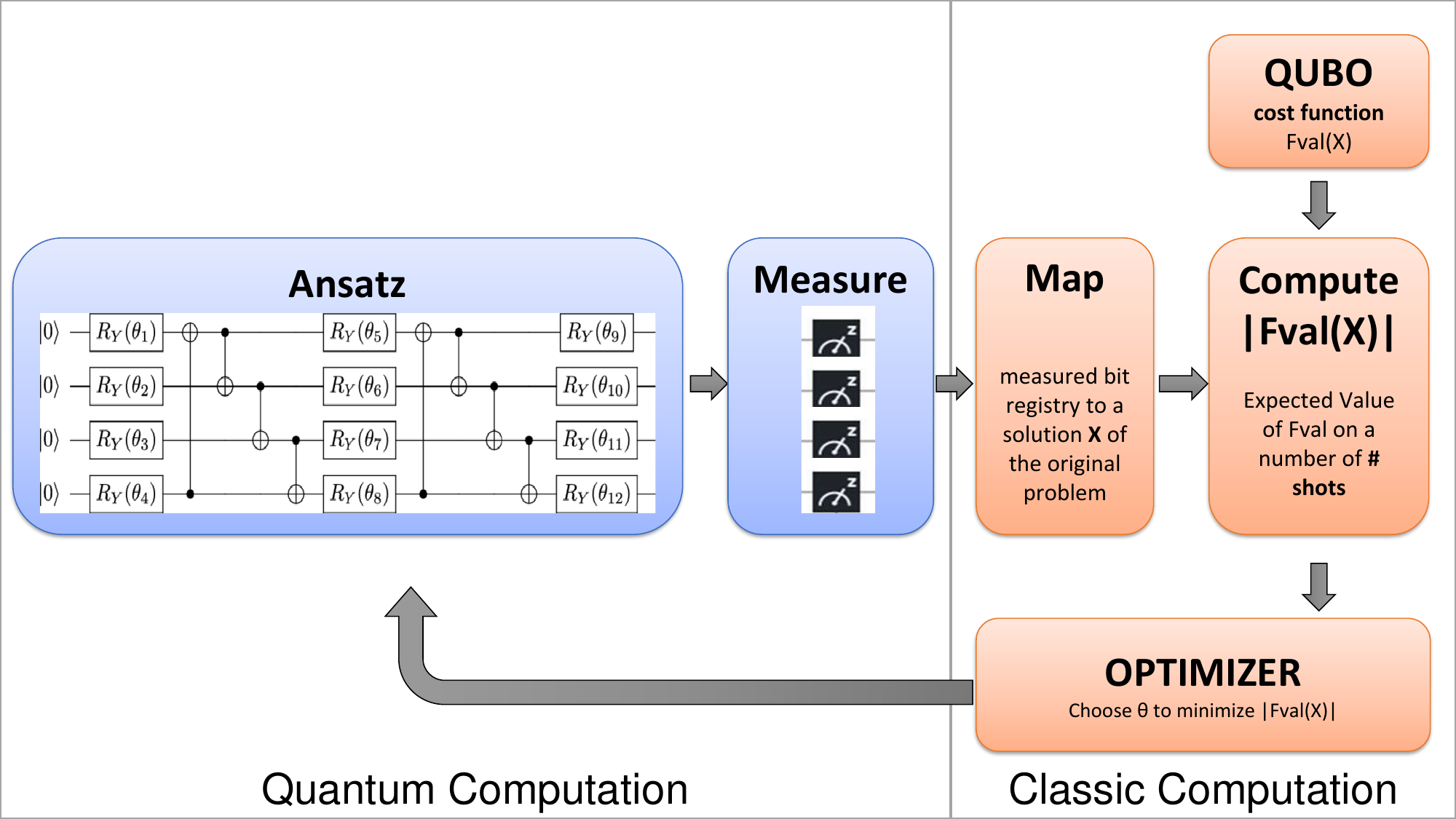}
\caption{Sketch of flow in the VQGAP algorithm.}
\label{fig:QGAP}
\end{figure}

At this point, all variables—both binary and integer—are defined, and the cost function given in Eq. \ref{eq:extObj} can be evaluated. 
Similarly to other hybrid quantum algorithms, a classical optimizer can find the parameters of the ansatz that minimize the expected value of the cost function.
The logical flow of the VQGAP algorithm is depicted in Figure \ref{fig:QGAP}.

With respect to the general VQE algorithm summarized in Section \ref{sec:vqe-intro}, the proposed strategy reduces the number of qubits required by the ansatz, as there is no need to model the binary slack variables introduced in Eq. (\ref{eq:ineq_to_eq}). Thus, the number of required qubits, Q, only depends on the assignment variables:
\begin{equation}
    Q=T \cdot (A+1)
\end{equation}

As a further consideration, the approach could also be exploited in problems that express budgets $B_j$ and weights $w_{ij}$ as positive real numbers instead of positive integer numbers, since there is no need to express the residuals $r_j$ as binary variables. 

\begin{figure}
    \centering
    \includegraphics[width=0.75\linewidth]{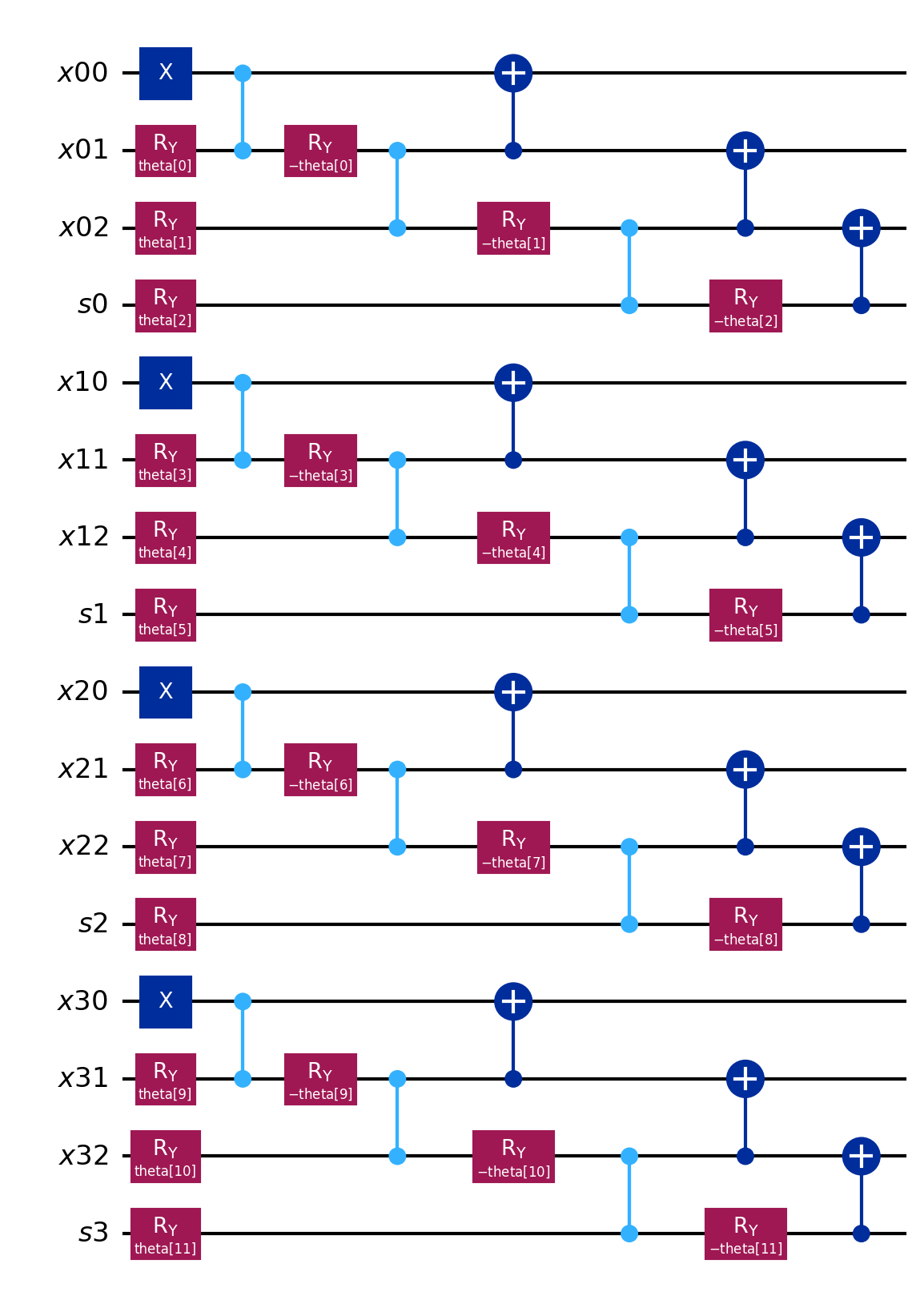}
    \caption{Reference ansatz for VQGAP}
    \label{fig:VQGAP-ansatz}
\end{figure}

A reference ansatz for VQGAP, for a problem with $T=3$ and $A=3$ is depicted in Figure \ref{fig:VQGAP-ansatz}. The ansatz is based on the one introduced in Section \ref{sec:vqe-intro}, and ensures the matching of the assignment constraints expressed by Eq. \ref{eq:vincoliProcesso}. 
Similar to the former introduced ansatz, the latter has $\Theta=T\cdot A$ tunable parameters.
The complexity in terms of number ($G_2$) and depth ($D_2$) of two-qubit gates is reduced, and is:

$\mathcal{O}(G_2) = 2T\cdot A$      

$\mathcal{O}(D_2) = 2A$

\subsection{VQGAPe approach}
As a second round of improvements, it is possible to decrease the number of required qubits by observing that, given the constraints expressed by Eq. (\ref{eq:vincoliProcesso}), each admissible solution satisfies the constraint that each task is assigned at most to one agent. Thus, for a general task $i \in \mathcal{T}$, the related variables $x_{ij}$ of a solution respecting such constraints should be all zeros  (no assignment) or one-up. By exploiting this observation, it is possible to produce an encoding/decoding of the problem that reduces the dimension of the solution space to be explored from $2^{T \cdot (A+1)}$, which considers all the possible values of the $ x_{ij}$ assignment variables, to $2^{T \cdot \log_2 (A+1)}$, which considers only the assignment values that satisfy the constraints expressed in Eq. \ref{eq:vincoliProcesso}. We call this second strategy VQGAPe (VQGAP encoded).

According to VQGAPe, given $T$ tasks and $A$ agents, we can design an ansatze defined on $Q=T*\lceil \log_2(A+1) \rceil$ qubits. For each task $i$ in $\mathcal{T}$, we design a set of encoded binary variables $e_{ik}$, with $1 \leq k < \lceil \log_2(A+1)\rceil$. Given a set of values for the $e_{ik}$ variables, we can compute the assignment variables $x_{ij}$ according to the following:


\begin{equation}
x_{ij}=\begin{cases}  
1,\ \ \text{if}\ \sum_{k} 2^{(k-1)} \cdot e_{ik} = j \\
0,\ \ \text{otherwise}
\end{cases}
\end{equation}

We, then, compute the residual values $r_j$ by applying the previously introduced Eq. (\ref{eq:residual-vqgap}).

\begin{figure}
    \centering
    \includegraphics[height=0.5\linewidth]{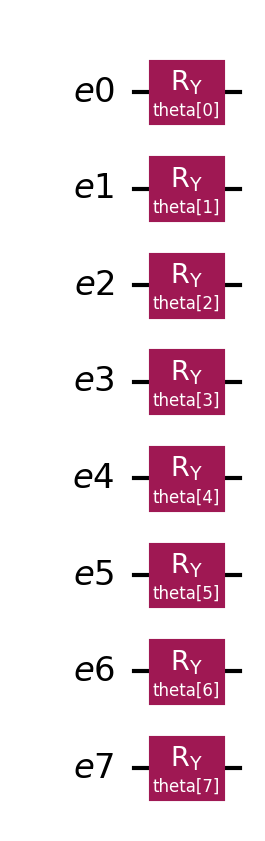}
    \caption{Ansatz VAQGAPe-RXL, consisting of a single layer of X-rotation on each qubit.}
    \label{fig:vqgape-rxlayer}
\end{figure}

\begin{figure}
    \centering
    \includegraphics[width=0.75\linewidth]{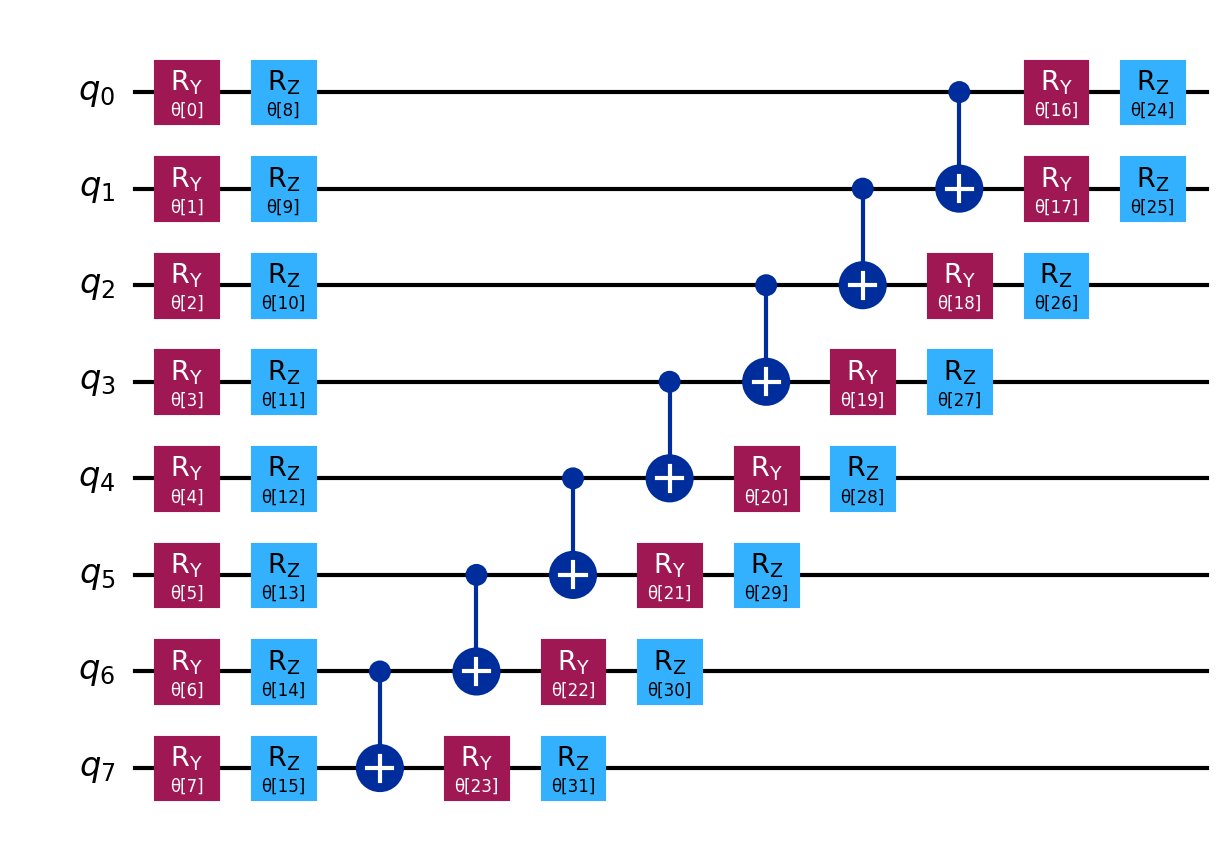}
    \caption{Ansatz VQGAPe-ESU2, an hardware efficient SU two-local parameterized circuit.}
    \label{fig:vqgape-esu2}
\end{figure}

As reference ansatze for VQGAPe, we propose the ones depicted in Figures \ref{fig:vqgape-rxlayer} and \ref{fig:vqgape-esu2}, and denoted as VAQGAPe-RXL and VQGAPe-ESU2. 

The VQGAPe-RXL ansatz (Fig. \ref{fig:vqgape-rxlayer}) consists of a single layer of parameterized rotations on the X axis. The number of tunable parameters is equal to the number of qubits, $\Theta=Q$ while $G_2=D_2=0$, since it does not encompass two-qubit gates. This is a very simple ansatz, that produces only separable solutions.

The VQGAPe-ESU2 ansatz (Fig. \ref{fig:vqgape-esu2}) is a hardware-efficient parameterized two-local circuit. It consists of an initial layer of parameterized single Ry and Rz rotations, followed by a series of control-not covering all the qubits in a linear pattern, and another layer of single Ry and Rx rotations. Vairations of this ansatz provides a number of replicas ($rep$) of the initial single rotation layer and the control-not layer.
The number of tunable parameters is equal to the number of qubits, $\Theta=Q \cdot (2+2\cdot rep)$ while $G_2=(Q-1)\cdot rep$ and $D_2=rep+(Q-1)$, since the replicated control-not layer interfoils each other.

\section{Results}
\label{sec:results}



In this section, we present a set of experimental results achieved by VQE, VQGAP and VQGAPe. The source code used to run the experiments and retrieve the results is available on a git repository\footnote{Code repository at \url{https://gitlab.com/qcc-icar-cnr/vqgap}}. 

The objective of the experiments was to assess the performance of VQE, VQGAP, and VQGAPe when solving general assignment problems of increasing sizes.

For this purpose, we executed the algorithms using the IBM Aer Statevector simulator, considering both an ideal quantum computer and a noisy one, by applying the noise model of a real quantum hardware, and specifically \textit{IBM Brisbane}.

We study noiseless simulations to forecast the potential performance of quantum hardware in the coming years. Moreover, noisy simulations are crucial for getting quick, approximate results on current quantum computers. 

The classic algorithm exploited to optimize the circuit parameters, for all of the experiments, was the Constrained Optimization by Linear Approximation (Cobyla) algorithm, in the implementation provided by scikit\-learn library. We set the number of shots (circuit execution and measurements) for each iteration to 4096.


We evaluated the following performance indices:
\begin{itemize}
\item the \textit{feasible solution probability}, $P_{feas}$, defined as the probability that the final measurement gives an admissible solution, i.e., a solution (optimal or non-optimal) that satisfies the constraints;
    \item the Coefficient of Performance\cite{montanezbarrera2023unbalanced}, for optimal and feasible solutions, respectively  $C_{best}$  and $C_{feas}$.  They are defined as the ratio between $P_{best [feas]}$ and the probability of obtaining an optimal [feasible] solution as a random guess: 
$C_{best [feas]}=P_{best[feas]}/(N_{best[feas]}\cdot 1/2^Q)$, where $N_{best[feas]}$ is the number of optimal [feasible] solutions. These indices help us to evaluate the ability of the quantum algorithms to amplify the probabilities of measuring useful basis states. 

    \item The absolute percentual error of the best solution found in the last iteration of the algorithm with respect to the optimal one, in terms of cost function.
    \item The absolute percentual error of the expected value of the cost function, considering the solutions explored in the last iteration.
\end{itemize}

All the reported results are the average of the results gathered by repeating the experiments 100 times.

We performed a set of experiments with two problem sizes. More specifically, we report  here the results obtained on the following GAP problems:

\begin{itemize}
\item GAP\_T4A3 - 4 Tasks, 3 Agents, $B_j\le 3$ for each agent $j$, encompassing 22 binary variables, according to equation \ref{eq:nQubits}
\item GAP\_T5A3 - 5 Tasks, 3 Agents, $B_j \le 3$ for each agent $j$, encompassing 26 binary variables.

\end{itemize}

We were not able to perform simulated runs on bigger problems due to the limitations of hardware and memory, since statevector simulation scales exponentially with respect to the number of qubits.

The rest of this section discusses the most interesting results of the experiments.

\subsection{Comparing VQE vs VQGAP}
\label{subsecComparingVQEvsVQGAP}

The first set of results shows a comparison between VQG and VQGAP, considering the previously introduced performance metrics.

\begin{figure}
    \centering
    \includegraphics[width=0.75\linewidth]{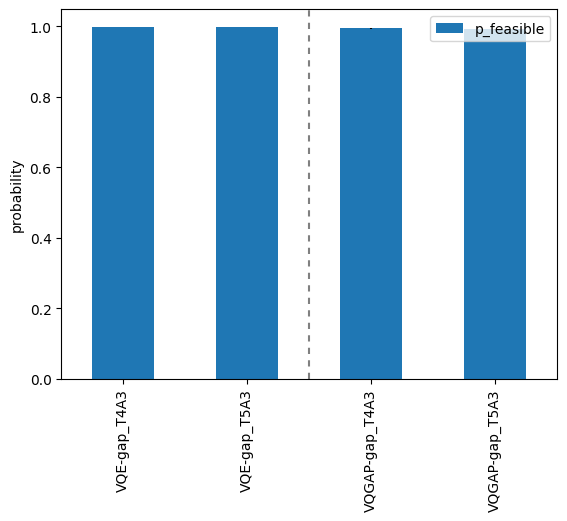}
    \caption{VQE vs VQGAP - Probability of finding a feasible solution for problems GAP\_T4A3 and GAP\_T5A3}
    \label{fig:vqevsvqgap-feasible}
\end{figure}

\begin{figure}
    \centering
    \includegraphics[width=0.75\linewidth]{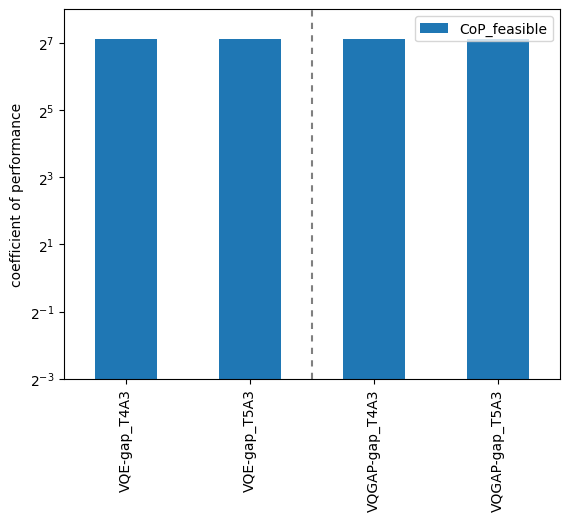}
    \caption{VQE vs VQGAP - Coefficient of Performance on problems GAP\_T4A3 and GAP\_T5A3}
    \label{fig:vqevsvqgap-cof}
\end{figure}

\begin{figure}
    \centering
    \includegraphics[width=0.75\linewidth]{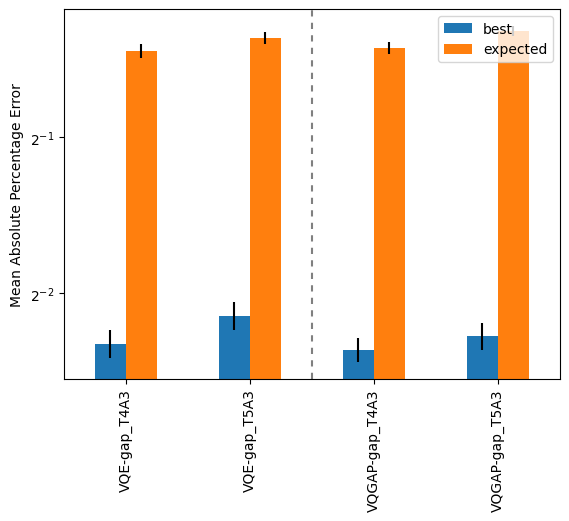}
    \caption{VQE vs VQGAP - Mean Absolute Percentage Error w.r.t. optimal solutions for problems GAP\_T4A3 and GAP\_T5A3}
    \label{fig:vqevsvqgap-mape}
\end{figure}

Figures \ref{fig:vqevsvqgap-feasible}, \ref{fig:vqevsvqgap-cof}, and \ref{fig:vqevsvqgap-mape} show, respectively, a comparison of the feasible solution probability $P_{feas}$, coefficient of performance, absolute percentual error obtained by exploiting VQE and VQGAP when solving the two problems GAP\_T4A3 and GAP\_T5A3, when executed on an ideal quantum computer.

From this first set of experiments, it is possible to observe that VQE and VQGAP show a very similar behavior, as expected, since the parametrized part of the exploited ansatze are very similar in both approaches. However, VQGAP is able to reach the performance of VQE, reducing the number of qubits and the depth of the circuits. This is due to the fact that, differently from VQE, VQGAP does not require explicitly expressing the binary variables related to the residual capacity of agents inside its ansatz, as mentioned  in Section \ref{sec:VQGAP}.

\subsection{VQGAP and VQGAPe}
\label{sec:vqgap-vqgape}

As a second set of experiments, we aim to compare the performance of the VQGAP approach, exploiting the reference ansatz shown in Figure \ref{fig:VQGAP-ansatz} with the VQGAPe approach, considering both the simple ansatz RXL and the hardware-efficient ansatz ESU2 reported, respectively, in Figures \ref{fig:vqgape-rxlayer} and \ref{fig:vqgape-esu2}.

Figures \ref{fig:vqgapvsvqgape-ideal-
    probabilities}, \ref{fig:vqgapvsvqgape-ideal-
    cop}, and \ref{fig:vqgapvsvqgape-ideal-
    mape}, report the results gathered considering an ideal, noiseless simulation. It is possible to observe that the compared approaches perform well in terms of feasible solution probability and coefficient of performance,  and  the VQGAPe-ESU2 shows the worst performance in terms of feasible solution probability. In contrast, when considering the absolute percentage error reported in Figure  \ref{fig:vqgapvsvqgape-ideal-
    mape}, it is possible to observe that the VQGAPe-ESU2 is able to find the best solutions among the approaches, even if, in the last iteration, it explores a solution space having an expected cost value worse than the others.

\begin{figure}
    \centering
    \includegraphics[width=0.75\linewidth]{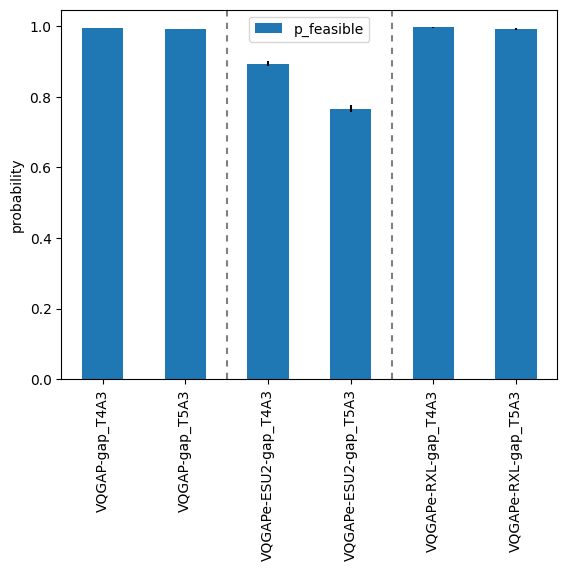}
    \caption{VQGAP and VQGAPe - Probability of finding a feasible solution for problems GAP\_T4A3 and GAP\_T5A3 (noiseless simulation)}
    \label{fig:vqgapvsvqgape-ideal-
    probabilities}
\end{figure}

\begin{figure}
    \centering
    \includegraphics[width=0.75\linewidth]{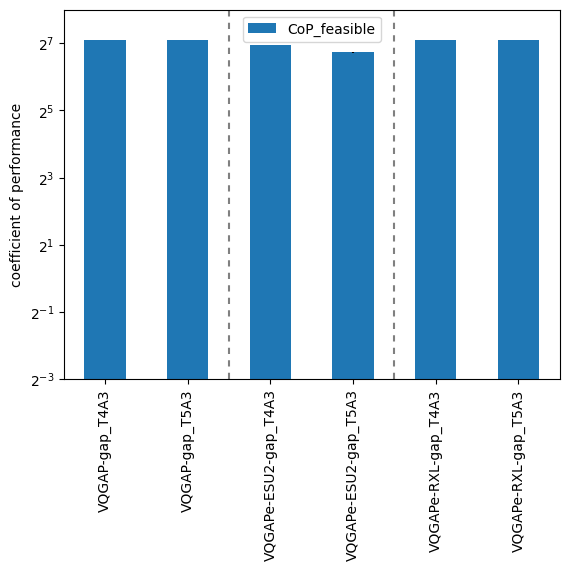}
    \caption{VQGAP and VQGAPe - Coefficient of Performance on problems GAP\_T4A3 and GAP\_T5A3 (noiseless simulation)}
    \label{fig:vqgapvsvqgape-ideal-
    cop}
\end{figure}

\begin{figure}
    \centering
    \includegraphics[width=0.75\linewidth]{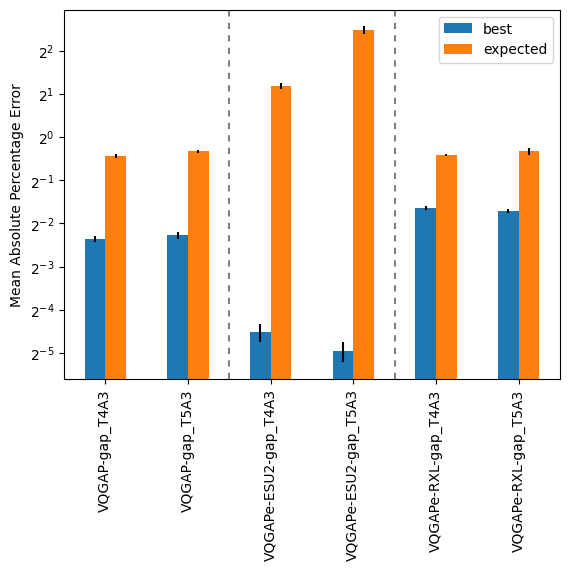}
    \caption{VQGAP and VQGAPe - Mean Absolute Percentage Error w.r.t. optimal solutions for problems GAP\_T4A3 and GAP\_T5A3 (noiseless simulation)}
    \label{fig:vqgapvsvqgape-ideal-
    mape}
\end{figure}

The same behavior is also observed when running the experiments considering noisy quantum hardware, with an interesting exception. As expected, introducing noise in the computation degrades the performance of the approaches in terms of probability of finding a feasible solution (Figure \ref{fig:vqgapvsvqgape-noisy-
    probabilities}), coefficient of performance (Figure \ref{fig:vqgapvsvqgape-noisy-
    cop} and expected percentage error of the cost function, even if it remains below 12\% in all the cases. However, it seems that the best solution found in the last iteration increases its quality in terms of percentage cost error w.r.t. the optimal solution. This is particularly significant for the VQGAPe-ESU2 approach, where the reduction is indeed important, as it has been shown to be capable of finding an optimal solution in all 100 runs for the problem GAP\_T4A3.

\begin{figure}
    \centering
    \includegraphics[width=0.75\linewidth]{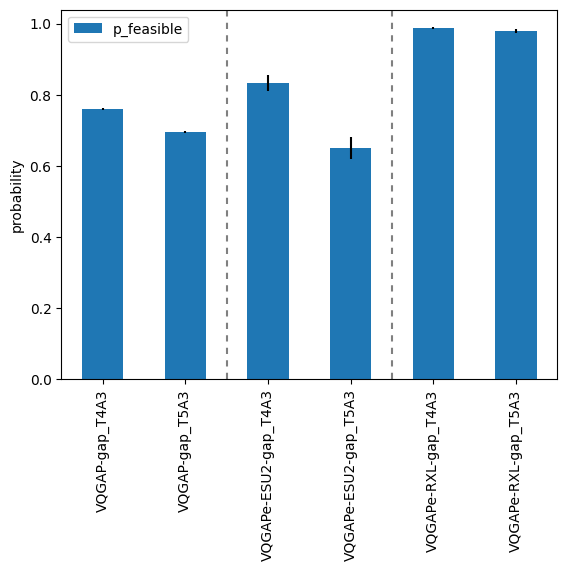}
    \caption{VQGAP and VQGAPe - Probability of finding a feasible solution for problems GAP\_T4A3 and GAP\_T5A3 (noisy simulation)}
    \label{fig:vqgapvsvqgape-noisy-
    probabilities}
\end{figure}

\begin{figure}
    \centering
    \includegraphics[width=0.75\linewidth]{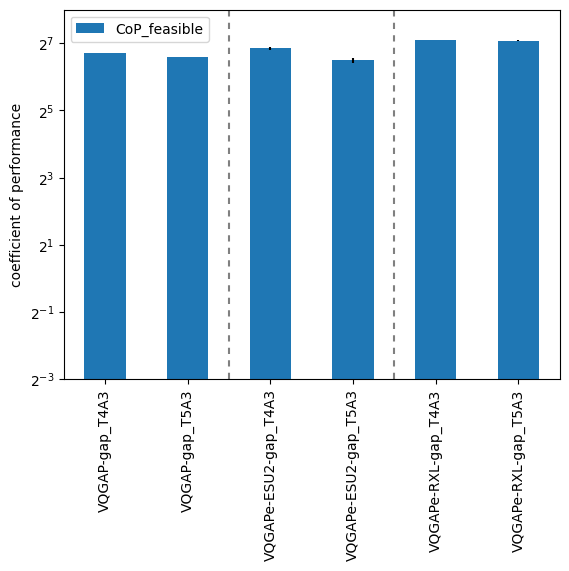}
    \caption{VQGAP and VQGAPe - Coefficient of Performance on problems GAP\_T4A3 and GAP\_T5A3 (noisy simulation)}
    \label{fig:vqgapvsvqgape-noisy-
    cop}
\end{figure}

\begin{figure}
    \centering
    \includegraphics[width=0.75\linewidth]{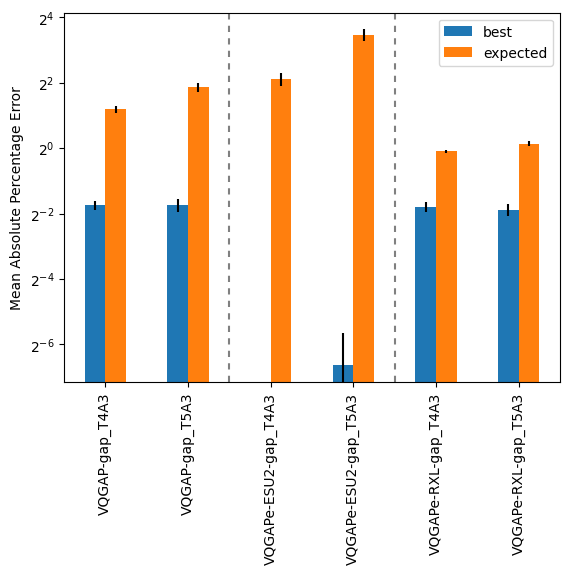}
    \caption{VQGAP and VQGAPe - Mean Absolute Percentage Error w.r.t. optimal solutions for problems GAP\_T4A3 and GAP\_T5A3 (noisy simulation)}
    \label{fig:vqgapvsvqgape-noisy-
    mape}
\end{figure}

As a final consideration, the reported comparison between VQGAP and VQGAPe shows that the latter approach can perform better than the former, while also significantly reducing the number of qubits required for its computation. As reported in Sections \ref{sec:VQGAP}, the number of qubits required to exploit VQGAPe scales logarithmically with respect to the number of agents of a considered GAP problem, while it scales linearly with respect to the number of agents in VQGAP (and VQE). This feature can be exploited to optimize the quantum resources to approach larger problems. 

\section{Conclusions and Future Work}
\label{sec:conclusions}

In this paper, we proposed a variation of the well-known VQE algorithm specifically tailored to approach the General Assignment Problem. The main innovation of the approach, namely VQGAP, consists in decoupling the ansatz and measurements from the computation of the expected value of the objective function, aiming at reducing the number of qubits required for solving a general GAP problem.

The preliminary results obtained here are, in this sense, very promising, since the approach seems to perform comparably to the VQE for GAP problems, while optimizing the required quantum resources. Considering the VQGAPe variant proposed in this work, the number of qubits required to tackle a GAP problem scales logarithmically with respect to the number of agents (while it scales linearly for VQE). Furthermore, the approach avoids the need to express slack variables related to the residual capacity of the agent in the ansatze.

Given this, a more in-depth investigation of the quality of the approach should be carried out, in particular by assessing it against a wider set of GAP problems and on real quantum hardware, and considering other aspects like scalability and overall computational time required for optimization. 

As another ongoing research, it is possible to investigate how the proposed approach could be generalized to optimization problems other than GAP.

\section*{Acknowledgments}
\footnotesize{
This work was partially funded by ICSC – Italian Research Center on High Performance Computing, Big Data and Quantum Computing, funded by European Union – NextGenerationEU, PUN: B93C22000620006, and by PNRR MUR project PE0000023-NQSTI, through the secondary project “ThAnQ”. This work was also supported by European Union - NextGenerationEU - National Recovery and Resilience Plan (Piano Nazionale di Ripresa e Resilienza, PNRR) - Project: “SoBigData.it - Strengthening the Italian
RI for Social Mining and Big Data Analytics” - Prot. IR0000013 - Avviso n. 3264 del 28/12/2021, and by European Union - NextGenerationEU - the Italian Ministry of University and Research, PRIN 2022 “INSIDER: INtelligent ServIce Deployment for advanced cloud-Edge integRation”, grant n. 2022WWSCRR, CUP H53D23003670006. JS acknowledges the contribution from PRIN (Progetti di Rilevante Interesse Nazionale) TURBIMECS - Turbulence in Mediterranean cyclonic events, grant n. 2022S3RSCT CUP H53D23001630006, CUP Master B53D23007500006. 
}

\bibliographystyle{IEEEtran}
\bibliography{bibliography}

\begin{thebibliography}{10}
\providecommand{\url}[1]{#1}
\csname url@samestyle\endcsname
\providecommand{\newblock}{\relax}
\providecommand{\bibinfo}[2]{#2}
\providecommand{\BIBentrySTDinterwordspacing}{\spaceskip=0pt\relax}
\providecommand{\BIBentryALTinterwordstretchfactor}{4}
\providecommand{\BIBentryALTinterwordspacing}{\spaceskip=\fontdimen2\font plus
\BIBentryALTinterwordstretchfactor\fontdimen3\font minus \fontdimen4\font\relax}
\providecommand{\BIBforeignlanguage}[2]{{%
\expandafter\ifx\csname l@#1\endcsname\relax
\typeout{** WARNING: IEEEtran.bst: No hyphenation pattern has been}%
\typeout{** loaded for the language `#1'. Using the pattern for}%
\typeout{** the default language instead.}%
\else
\language=\csname l@#1\endcsname
\fi
#2}}
\providecommand{\BIBdecl}{\relax}
\BIBdecl

\bibitem{VQA}
M.~Cerezo, A.~Arrasmith, R.~Babbush, S.~C. Benjamin, S.~Endo, K.~Fujii, J.~R. McClean, K.~Mitarai, X.~Yuan, L.~Cincio \emph{et~al.}, ``Variational quantum algorithms,'' \emph{Nature Reviews Physics}, vol.~3, no.~9, 2021.

\bibitem{VQA-IEEE-TQE}
J.~Wurtz and P.~J. Love, ``Classically optimal variational quantum algorithms,'' \emph{IEEE Transactions on Quantum Engineering}, vol.~2, pp. 1--7, 2021.

\bibitem{maxcut}
Z.~Wang, S.~Hadfield, Z.~Jiang, and E.~G. Rieffel, ``Quantum approximate optimization algorithm for maxcut: A fermionic view,'' \emph{Physical Review A}, vol.~97, no.~2, p. 022304, 2018.

\bibitem{maxsat}
K.~Marwaha and S.~Hadfield, ``Bounds on approximating max $ k $ xor with quantum and classical local algorithms,'' \emph{Quantum}, vol.~6, p. 757, 2022.

\bibitem{routing-IEEE-TQE}
S.~Harwood, C.~Gambella, D.~Trenev, A.~Simonetto, D.~Bernal, and D.~Greenberg, ``Formulating and solving routing problems on quantum computers,'' \emph{IEEE Transactions on Quantum Engineering}, vol.~2, pp. 1--17, 2021.

\bibitem{IsingNP}
\BIBentryALTinterwordspacing
A.~Lucas, ``Ising formulations of many np problems,'' \emph{Frontiers in Physics}, vol.~2, 2014. [Online]. Available: \url{https://www.frontiersin.org/article/10.3389/fphy.2014.00005}
\BIBentrySTDinterwordspacing

\bibitem{preskillNisq}
J.~Preskill, ``Quantum computing in the nisq era and beyond,'' \emph{Quantum}, vol.~2, p.~79, 2018.

\bibitem{NISQ-IEEE-TQE}
E.~Pelofske, A.~Bärtschi, and S.~Eidenbenz, ``Quantum volume in practice: What users can expect from nisq devices,'' \emph{IEEE Transactions on Quantum Engineering}, vol.~3, pp. 1--19, 2022.

\bibitem{biamonte2017}
J.~Biamonte, P.~Wittek, N.~Pancotti, P.~Rebentrost, N.~Wiebe, and S.~Lloyd, ``Quantum machine learning,'' \emph{Nature}, vol. 549, no. 7671, 2017.

\bibitem{dunjko2016}
V.~Dunjko, J.~M. Taylor, and H.~J. Briegel, ``Quantum-enhanced machine learning,'' \emph{Physical review letters}, vol. 117, no.~13, p. 130501, 2016.

\bibitem{QuantumOptimization2018}
N.~Moll, P.~Barkoutsos, L.~S. Bishop, J.~M. Chow, A.~Cross, D.~J. Egger, S.~Filipp, A.~Fuhrer, J.~M. Gambetta, M.~Ganzhorn, A.~Kandala, A.~Mezzacapo, P.~Müller, W.~Riess, G.~Salis, J.~Smolin, I.~Tavernelli, and K.~Temme, ``Quantum optimization using variational algorithms on near-term quantum devices,'' \emph{Quantum Science and Technology}, vol.~3, no.~3, p. 030503, June 2018.

\bibitem{IEEESmartGrid}
C.~Mastroianni, F.~Plastina, L.~Scarcello, J.~Settino, and A.~Vinci, ``Assessing quantum computing performance for energy optimization in a prosumer community,'' \emph{IEEE Transactions on Smart Grid}, vol.~15, no.~1, pp. 444--456, 2024.

\bibitem{QuantumEdgeCloud}
C.~Mastroianni, F.~Plastina, J.~Settino, and A.~Vinci, ``Variational quantum algorithms for the allocation of resources in a cloud/edge architecture,'' \emph{IEEE Transactions on Quantum Engineering}, vol.~5, pp. 1--18, 2024.

\bibitem{QAOA}
E.~Farhi, J.~Goldstone, and S.~Gutmann, ``A quantum approximate optimization algorithm,'' 2014.

\bibitem{VQE}
A.~Peruzzo, J.~McClean, P.~Shadbolt, M.-H. Yung, X.-Q. Zhou, P.~J. Love, A.~Aspuru-Guzik, and J.~L. O’brien, ``A variational eigenvalue solver on a photonic quantum processor,'' \emph{Nature communications}, vol.~5, no.~1, p. 4213, 2014.

\bibitem{ising2000}
B.~A. Cipra, ``The ising model is np-complete,'' \emph{SIAM News}, vol.~33, no.~6, 2000.

\bibitem{IEEE-TQE-2024}
C.~Mastroianni, F.~Plastina, J.~Settino, and A.~Vinci, ``Variational quantum algorithms for the allocation of resources in a cloud/edge architecture,'' \emph{IEEE Transactions on Quantum Engineering}, vol.~5, pp. 1--18, 2024.

\bibitem{montanezbarrera2023unbalanced}
\BIBentryALTinterwordspacing
A.~Montanez-Barrera, D.~Willsch, A.~Maldonado-Romo, and K.~Michielsen, ``Unbalanced penalization: A new approach to encode inequality constraints of combinatorial problems for quantum optimization algorithms,'' 2023. [Online]. Available: \url{https://arxiv.org/abs/2211.13914}
\BIBentrySTDinterwordspacing

\end{thebibliography}

\vfill


\end{document}